# The "optical" version of the barn-pole problem

**Eric Baird**  (eric_baird@compuserve.com)

We present diagrams and simple calculations for the apparent (i.e. photographable) length of a moving ruler skimming the observer's position, under three different classes of model. Special relativity's predictions in this particular situation are the root-product average of the two more basic first-order predictions generated by simple "propagation timelag" arguments. We find that special relativity can legally predict either a photographable Lorentz contraction or a photographable Lorentz expansion in the "centred" ruler, depending on whether our camera is at the ruler's "apparent" or "official" centre.

## 1. Introduction

One of special relativity's most contentious issues has been whether the length contraction effect originally proposed by Fitzgerald and Lorentz [1] should be considered to be "real" or not in Einstein's model [2] (some authors argue that this question has no easy answer, e.g. [3] pp.76-77). In 1959 Terrell and Penrose independently identified what seemed to be a fifty-year misinterpretation of special theory's predictions for photographable lengths [4][5], prompting a flurry of related papers [6]-[20], and inspiring the famous "barn/pole" or "polevaulter" problem [7] and a range of similar paradoxes or pseudo-paradoxes [8][9].

We take a fresh look at the "polevaulter" problem by setting aside the usual arguments about calibrated coordinate systems, and instead using the distance information that will be recorded in a photographic image according to special relativity and other models.

We find that, for a central "barn" observer, different definitions of transverse motion can lead to photographs being taken at different times, so that we obtain distinctly different images in which the overall length of the "transverse-moving" pole is seen to be either contracted or elongated. In each case, special relativity's prediction is the "root product" average of the (non-paradoxical!) lengths that we would expect to see according to basic signal-propagation arguments [21][22].

## 2. Non-transverse Doppler effects

The three fundamental radial Doppler equations for an object receding at *v* m/s are

$$freq' / freq = (c-v) / c \qquad \dots (1)$$

$$freq' / freq = \sqrt{(c-v)/(c+v)} \qquad \dots (2)$$

$$freq' / freq = c / (c+v) \qquad \dots (3)$$

, where **(1)** and **(3)** are the basic propagation shift predictions associated with a "flat" medium stationary in the emitter and observer frames, and **(2)** is special relativity's intermediate "relativistic Doppler" formula (*v* becomes negative for approach velocities).

## 3. Simple optical ruler-effects

### Recession or approach

If a luminous ruler approaches or recedes from the observer, its apparent length in the direction of motion decreases or increases by the same ratio as the received frequency of its (Doppler shifted) light ("spatial equivalent of the Doppler effect" [23], [24][25][26]). The Doppler equations **(1)-(3)** dictate apparent lengths as well as apparent frequencies [27].

### Transverse motion (special case)

While the ruler is passing the observer, the receding part of the ruler will appear to be contracted and the approaching part will appear to be elongated, and the total apparent length will then simply be the sum of these two apparent lengths.

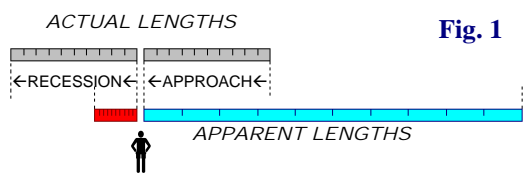

**Fig. 1**

When the ruler is centred on the observer's position, this calculation gives us a model's prediction for the apparent length of a passing ruler for the special case of zero separation from the observer.

### Defining "centre"

Although this method gives us an easy way of calculating a special-case transverse result from non-transverse Doppler relationships, we find that the outcome is sensitive to the precise moment at which the observer chooses to look at (or photograph) the moving ruler. Our "transverse" prediction has different values depending on whether we choose to look at the object at the moment when it is *thought* to be centred on the observer's position, or at the moment when it *seems* to be centred on the observer's position.  In other words, we get different "transverse" predictions depending on whether the observer coincides with the ruler's "factory-marked" mid-point or its "apparent" midpoint.

## 4. Factory-marked centre

For our first calculation, a ruler of nominal unit length is photographed at the exact moment that its factory-marked mid-point is adjacent to the camera.

If we add the apparent lengths of the approaching and receding halves of the ruler, our three alternative Doppler equations give us predictions for the total apparent length of:

$$0.5 \frac{c-v}{c} + 0.5 \frac{c-(-v)}{c} = 1 \quad \text{… (i)}$$

$$0.5 \sqrt{\frac{c-v}{c+v}} + 0.5 \sqrt{\frac{c-(-v)}{c+(-v)}} = \frac{1}{\sqrt{1-v^2/c^2}} \quad \text{… (ii)}$$

$$0.5 \frac{c}{c+v} + 0.5 \frac{c}{c+(-v)} = \frac{1}{1-v^2/c^2} \quad \text{… (iii)}$$

## 5. Apparent centre

If we want to instead engineer a *visible* Lorentz contraction, we can choose to take our photograph at a slightly later time, when more of the passing ruler's structure will be receding than approaching.

At the moment when the ruler's receding and approaching parts ruler are *seen* to have equal lengths, the total apparent lengths associated with each Doppler equation will be:

$$len' / len = 1 - v^2/c^2 \quad \text{… (i)}$$

$$len' / len = \sqrt{1 - v^2/c^2} \quad \text{… (ii)}$$

$$len' / len = 1 \quad \text{… (iii)}$$

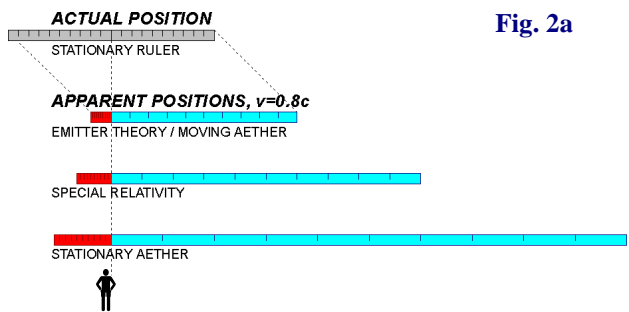

**Fig. 2a**

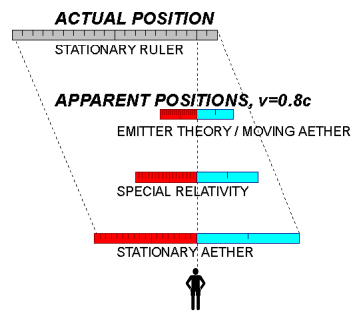

**Fig. 2b**

In each case, there is an offset between the object's apparent mean position, $[(x_{FRONT}+x_{REAR})/2]$, and the its nominal centre of gravity, $[(restlength_{RECEDING} = restlength_{APPROACHING})]$.

With equation **(2)** (special relativity), our centred ruler is seen to have an overall length that is Lorentz-dilated rather than Lorentz contracted [26][28].

**Note:** Under special relativity, an approaching or receding object is seen to be Lorentz-contracted *compared to tour expectation of its apparent length if we assumed that light "really" propagated throughout space at fixed speed in the observer's frame*.
This "general" contraction effect is interpretative and depends on the belief system of the observer, since the same SR-compatible images also show a Lorentz *elongation* relative to our expectations for a simple aether fixed in the *object's* frame. The decision to describe an SR-photographed object as being "really" contracted or "really" elongated is without physical consequence.

Similar arguments apply to time-dilation effects – an (SR) photographed image can be interpreted as including either a redshift component ("time dilation") or a blueshift component ("Lorentz boost"), depending on whether we choose to compensate for the assumed first-order Doppler effects associated with equation **(1)** or **(3)**.

However, under special relativity there is not generally a viewable contraction in length relative to the object's *rest* length, unless the stationary-aether prediction for apparent length happens to be the same as the rest length value, as it is in this particular special case.

## 6. Barn-pole sequence (special relativity version)

Figure 3 (below) illustrates the actual pole positions (faint outline) and the apparent pole positions (solid outline) that are seen by a centrally-placed barn observer in special relativity's "barn-pole" problem [7]-[9] as the pole passes by the observer. The diagram shows a pole and barn both nominally 200m long, with factory markings at 10m intervals. The pole is moving from right to left, and the relative velocity is $v=0.8c$, giving a Lorentz factor of $\gamma=0.6$, and recession and approach length-change ratios (from equation (2)) of 1:3 and 3:1.

We have selected four successive pole positions, where the central barn observer's position coincides with **a)** the pole front, **b)** the pole's "nominal" centre, **c)** the pole's "apparent" centre, and **d)** the pole's rear.

If we calculate the apparent lengths using only the assumption of flat spacetime and propagation fixed to the observer's frame (3), we get values of:
**a)** 1000m, **b)** 555.5'm, **c)** 200m, and **d)** 111.1'm.
Special relativity's predictions are shorter than these values by the Lorentz ratio.

Equivalent calculations from the assumption of light-propagation fixed in the *pole*'s frame (1) give:
**a)** 360m, **b)** 200m, **c)** 72m, and **d)** 40m.
Special relativity's predictions are the root product average of these two sets of first-order calculations.

Under special relativity, if the barn and pole have the same nominal rest length and the relative velocity exceeds $0.6c$, the central barn observer will not see the front of the pole reaching the rear of the barn until the entire length of the pole is already receding from the observer.

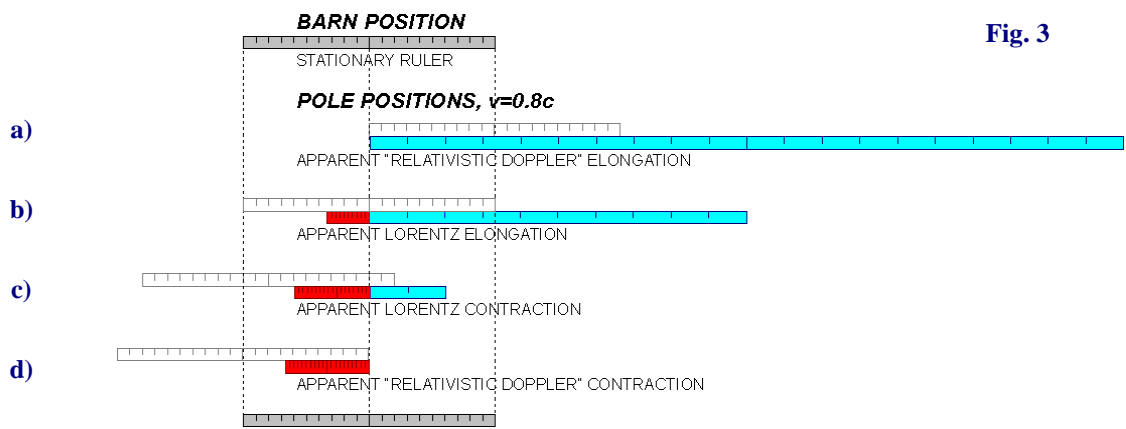

**Fig. 3**

The apparent lengths of the pole in these positions (as seen by the central barn observer) are then:

**a)** The barn observer sees the approaching pole to be **600m** long.

**b)** The barn observer is adjacent to the factory-marked centre of the pole, and sees the receding 100m to be reduced in length to 33.3'm, and the approaching 100m to be dilated in length to 300m. The total apparent length is **333.3'm**, which is a Lorentz elongation (333.3'm = 200m / ($\gamma=0.6$) ).

**c)** 180m of pole has already passed the observer, and only 20m is still approaching. The total apparent length is now (180/3) + (20×3) = 60 + 60 = **120m**. The pole appears to be centred on the observer's position, and appears to have an overall Lorentz length contraction (120m = 200m × ($\gamma=0.6$) ).

**d)** The final 20m of pole has passed, and the total length of the receding pole now appears to be 200/3 = **66.6'm**.

It is important to remember that in **c)**, where the pole is seen to occupy a Lorentz-contracted volume of barn space (and could be seen to fit into a correspondingly smaller barn), this apparent contraction effect does not apply equally to all parts of the pole structure. The object's centre-of-gravity will appear displaced towards the receding side, and since more of the pole's structure is receding from the observer than approaching, the pole's signals will (in one sense) also have an "average" redshift.

To obtain this "visible" transverse Lorentz contraction result, we have to define the "transverse-moving centre" of our object as being the mid-point of object's two apparent end coordinates. For an apparently-centred ruler of total length **L+L**, the offset between the ruler's "apparent" and "official" centres is

$$\mathit{offset} = \mathbf{L}\, v/c$$

This relationship (where *offset* is a ruler-marked quantity) holds for all three Doppler formulae.

***Example:*** *at v=0.8c, a 200m ruler will be seen to be centred when the observer is directly alongside the ruler's "180m" mark.*

## 7. Conclusions

In the limited situations described here, where our object subtends a 180° angle with the observer …

1. … We should expect to see apparent ("photographable") length-contraction and expansion effects in passing objects, as basic first-order signal-propagation artifacts.

2. … Special relativity's predictions for apparent lengths are the root product average of the two corresponding first-order predictions.

3. … A single SR-compatible photograph can be *interpreted* as demonstrating the existence of Lorentz contraction or Lorentz expansion, depending on whether we decide to compensate for presumed first-order length-change effects associated with **(1)** or **(3)**.

4. … The SR photograph can also show a *directly-photographable* overall Lorentz contraction or expansion relative to the object's rest length, depending on whether the camera shutter is released when the object appears to be, or is deduced to be, centred on the camera's position.

Although the first three points are already documented, the fourth (sensitivity of the outcome to our definition of "transverse" motion) is not so widely appreciated.

We would argue that the special theory's results should only be presented as paradoxical if they generate numerically-conflicting predictions for the same *directly-observed* (uninterpreted) data. Since apparent first-order length changes are not deemed paradoxical (they are also, after all, the conventional Doppler relationships for audio signals), special relativity's average of these two first-order predictions should also not be considered to be paradoxical (or radically different to earlier models).

Much of the confusion about special relativity's validity can be blamed on confusion over the correct meaning of "transverse" motion and on the common use of interpretative coordinate-system arguments (which can introduce additional conflicting assumptions and beliefs about the propagation of light) rather than on the theory's actual, verifiable physical predictions.